\documentclass[aps,floatfix,twocolumn,nofootinbib,amsmath]{revtex4}
\usepackage{epsfig}
\begin{document}
\def\mod{{\rm mod}}
\def\tr{{\rm tr}}
\def\rhob{{\boldsymbol\rho}}
\def\sigmab{{\boldsymbol\sigma}}
\def\be{\begin{equation}}
\def\ee{\end{equation}}
\def\Do{{\Delta^{(3)}_o}}
\def\De{{\Delta^{(3)}_e}}
\def\lb{\langle}
\def\rb{\rangle}

\title{
Whence the odd-even staggering in nuclear binding?
}
\author{  
W.A.~Friedman and G.F.~Bertsch}
\affiliation{
Department of Physics and Institute of Nuclear Theory,
Box 351560\\
University of Washington
Seattle, WA 98915 USA\\
}
\begin{abstract}
We explore the systematics of odd-even mass staggering with a view to 
identifying the physical mechanisms responsible. The BCS pairing and
mean field contributions have $A$- and number parity dependencies
which can help disentangle the different contributions.
This motivates the two-term parametrization $c_1 + c_2/A$ as
a theoretically based alternative to the inverse power form traditionally
used to fit odd-even mass differences.
Assuming that the $A$-dependence of the BCS pairing is weak, we find that
mean-field contributions are dominant below mass number $A\sim 40$ while
BCS pairing dominates in heavier nuclei.

\end{abstract}
\date{INT-PUB 08-56}
\maketitle
\section{Introduction} 
The observed odd-even mass staggering
has been extensively explored in the literature\cite{sa98,ru99,do01,hi02,be03}
The effect has generally been associated with the  
pairing gap $\Delta$, as suggested by BCS theory.
When examined over the  full table  of measured 
nuclides \cite{au03}, the values 
for the staggering display a great deal of  scatter. 
However,  a slow trend with nuclear mass number 
$A$ is also observed.  This trend   has been 
extensively analyzed in the
literature by fitting data to an inverse power of 
$A$.
Traditionally this has been  $1/A^{1/2}$, but 
other powers are used as well\cite{hi02,olo08,mol92}.
For example, ref. \cite{hi02} advocates an expansion in powers
of $A^{-1/3}$, based on the finite-size analysis in ref. \cite{fa03}. 
There is no theoretical justification for a 
single-term fractional-power dependence, however.    

It has been suggested that,  
in addition to the anticipated  BCS type of pairing,
 
other effects contribute significantly to the staggering \cite{sa98,do01,be03}. 
One would anticipate 
that each contribution would have its  own  characteristic dependence on
$A$.
Consequently the staggering should
follow some combination of these. 
The tradition of representing  the net 
effect by a form with a single fractional exponent obscures the complexity of  the  
effect. We propose here a different form that may offer a better opportunity
to identify the various contributing effects.

There are a number of difference formulas that may be used as measures for the
staggering\cite{BM,mol92,ben00}.  In this work we use the measure conventionally
called  $\Delta^{(3)}$ \cite{sa98,do01,hi02}. It involves 
the difference between the two separation energies of adjacent nuclei.
Here the removal of one type of nucleon,
i.e., proton or neutron, is considered, while the number of the opposite 
isospin is kept the same. We believe it 
is the most direct measure of staggering and is the
most convenient quantity for theoretical study. 
One defines the 
{\it neutron 
OES} as
\be
\label{d3o}
\Delta^{(3)}_{on}(N,Z)) =-(-)^{\pi_N}{1\over 2}\left (S_n(N+1,Z) - S_n(N,Z)\right)
\ee
Here $S_n$ is the separation energy, $S_n(N)=B(N)-B(N-1)$ and $B$ the binding energy.
The factor depending on the number parity $\pi_N$ is chosen so that the OES
centered on even and odd neutron numbers $N$ will both be positive
for normal attractive pairing.
The proton OES are defined similarly and we designate them by
$\Delta^{(3)}_{op}$ and $\Delta^{(3)}_{ep}$.

The 4-point measure used by Bohr and Mottelson \cite[Eqs. 2-92, 2-93]{BM}
is the average of adjacent even and odd values of
$\Delta^{(3)}$. We will also find it convenient to examine the average
and differences of our parametrized OES, but that does not require
contiguous data.
It should also be noted that, because the $\Delta^{(3)}$ values  represent
second differences of the binding energies, they may  be 
non-zero even without any odd-even staggering. 
  
The various OES offer interesting possibilities for examining
the various contributions. First, the even and odd OES 
have different contributions that can possibly be
distinguished.  Second, the two isospins (neutron and proton) can
be compared to give some indication of the role of the Coulomb
and other isospin dependent effects.

\section{Mechanisms}
There are a several mechanisms that
contribute to OES in framework of mean field theory.
We consider
three of them: single-particle energies, BCS pairing, and orbital
interaction energies.
We will assume that orbitals are two-fold degenerate between time-reversed
partner states to estimate single-particle and diagonal interaction
energies.
A summary of the following discussion is provided 
in Table I below,  which lists separately  the  contributions  pertinent to
$\Delta^{(3)}_o$  and  $\Delta^{(3)}_e$.   There are several kinds of
orbital interaction that affect the systematics of the OES.
Most important are the spin- and symmetry-dependent
interactions that are classified under ``time-odd fields'' in the
density functional formulation of mean-field theory.  But also the
Coulomb interaction and the ordinary isospin-spin dependent interactions
play a role when examining the differences of $\Delta^{(3)}$ measures.
Another mechanism that has been suggested \cite{sa98,do01} involves
the static polarization of the core by the valence orbitals.  However,
it has no obvious parametric
dependence on $A$, and we must assume that its contribution is small to
carry out the present analysis.

\begin{table}[h!]
\caption{Contributions to odd-even staggering of nuclear binding energies,
and their assumed $A$-dependence.}
\begin{tabular}{|c|cc|c|}
\colrule
Mechanism & $\Delta^{(3)}_o$  & $\Delta^{(3)}_e$ & $A$-dependence\\
\colrule
Single-particle & 0  & $ (e_i -e_{i-i})/2$ & $1/A$\\
BCS correlation& $ \Delta_{BCS} $ & $ \Delta_{BCS} $ &  const. \\
time-odd    &  $-v_{i \bar i,i\bar i} /2 $  &  
     $ \bar v -v_{i \bar i,i\bar i} /2 $ &  $1/A$\\     
\colrule
\end{tabular}
\end{table}

\subsection{Single-particle energies}
  This mechanism is discussed in many places, e.g., ref. \cite{sa98}, and we
repeat the argument for completeness.
The binding energy 
associated with the single-particle 
Hamiltonian is the negative of the eigenvalue $e_i$. The neutron (or proton) 
orbitals are doubly degenerate implying that the contribution from the
two separation energies odd OES.  On the other hand,
the two separation energies in the even OES come from different orbitals.
The resulting OES are 
\be
\label{sp}
\Delta^{(3)}_o(s.p.)= 0 ;\,\,\,\,\,  \Delta^{(3)}_e(s.p.)=  (e_i -e_{i-1})/2. 
\ee
For brevity  we will refer to this as the ``single-particle" contribution.  
The average difference in single-particle energies can be
estimated in the Fermi gas model as \cite{BM}
\be
\label{fg}
e_i -e_{i-1} \approx 4 e_F/3 N \approx 93/A,\,{\rm MeV}.
\ee
In the last approximate equality we have estimated the level spacing
using $e_F\approx 35$ MeV and $A\approx 2 N$.
\subsection{BCS correlation energy}
  To avoid confusing the BCS correlation energy with the mean-field
energies, let us define it  as the additional energy associated with an
interaction of the form
\be
\label{vpair}
\sum'_{i\ne j} v_{i \bar i,j \bar j}a^\dagger_i a^\dagger_{\bar i} a_{\bar j}
a_j.
\ee
The orbital $\bar i$ is the time-reversed partner of orbital $i$.
The prime on the summation indicates 
that a sum over  $i$  takes only one member of the
partners $(i,\bar i)$.  The restriction to $i\ne j$ ensures that there is 
no mean-field contribution.  
In the large number limit, the BCS energies give the familiar result,
$\Delta^{(3)}_{o,e} = \Delta_{BCS}$, where $\Delta_{BCS}$ is the pairing
condensate energy for an orbital at the Fermi surface. We will refer to
this contribution  as the ``pairing" contribution.  

The $A$-dependence of the BCS pairing is an open question at present. 
An analytic treatment can be carried out\cite{fa03} following the 
derivation of the BCS formula,
\be
\label{BCSgap}
\Delta(BCS) =  2Se^{-1/(G\rho(\mu))}.
\ee
Here  $S$ is the half-width of the  truncation zone, $G$ is an average 
pairing interaction matrix element and $\rho(\mu)$ is the  
density of levels at the Fermi-energy, $\mu$.  For large $A$, we have
$G\sim 1/A$ and $\rho \sim A$ giving a constant $\Delta$.  At the next
order in a 
finite-size expansion  both $G$ and $\rho$ add terms varying as 
$A^{-1/3}$.  This gives an overall
$A$-dependence  $\Delta(BCS) \sim  \exp(c/A^{1/3}) $ \cite{fa03}.  
However, the
finite range of the nuclear interaction as well as the induced pairing
introduces additional $A$ dependencies that are hard to estimate.
In a recent global study \cite{be08} it
was found that the $A$ dependence of the average OES is quite weak,
taking pairing interactions with density dependence as commonly used.
We shall therefore make the simplest assumption here, that the
BCS correlation energy contribution is independent of $A$.
Finally, we note that although Eq. (\ref{sp}) was derived only for the Hartree-Fock
limit, it remains approximately valid in the presence of the pairing
Hamiltonian Eq. (\ref{vpair}).

\subsection{Time-odd fields}  
  A semi-analytic treatment of this mechanism is given in ref. \cite{du01}.
There it is called a polarization effect, but in fact it
is present in the diagonal orbital interaction.  
The expressions for the effect
are quite different
depending on whether one is using a density functional formalism or a 
formalism based on interaction matrix elements between orbitals.  The first
formalism has unphysical self-energies that are canceled by the exchange
term in the interaction.  In the second formalism self-interactions do
not appear and one does not make use of single-particle fields and their
transformation properties under time reversal.  For simplicity in the  following 
we  follow the approach 
based on interaction matrix elements.
In this formulation,
we count the interactions as particles are added to the system, always 
filling both orbitals of a Kramers pair if possible.  Suppose we start
with an even-even system with $A$ nucleons and add
another neutron in orbital $i$.  The additional interaction
energy is
$$ 
V_i = {1\over 2}\sum_j \left(v_{ij,ij} + v_{i\bar j,i\bar j}\right)=A \bar v.
$$
Here the $j$ summation runs over all occupied orbitals except $i$ and we
have defined an average interaction energy $\bar v$ in the last
equality. 
We next add another particle to the system in 
orbital $\bar i$, giving an additional interaction energy,
$ V_i  + v_{i\bar i,i\bar i}$.  Then 
$\Delta^{(3)}_o$ is given by half the difference,
\be
\label{to-odd}
\Delta^{(3)}_o  =  -v_{i\bar i,i\bar i}/2
\ee
Adding a third particle gives an additional interaction energy of 
$(A+2) \bar v$.  This can be used to obtain the following estimate for
$\Delta^{(3)}_e$
\be
\label{to-even}
\Delta^{(3)}_e =  \bar v - v_{i\bar i,i\bar i}/2.
\ee
For the particle-particle 
interaction matrix elements  associated with the time-odd  contribution we 
use the  same argument as with the 
pairing matrix elements $G$. This leads to the expectation that the
average matrix elements embodied in the energies $\bar v$ and $v_{i \bar i,i \bar i}$
also would have
an $A^{-1}$ dependence.  The formulas Eq. (\ref{to-odd}) and (\ref{to-even}) for the time-odd
contribution were derived assuming
that the orbital occupancies are zero or one, but the formulas
remain valid in the presence of pairing in the nonblocked orbitals, 
as was the case for the single-particle contribution. 
 
\section {Examining the experimental data}
We now examine the experimental data on the OES, fitting it to 
various parametrizations.  The data sets are defined with the
same criteria used in ref. \cite{be08}, using the Audi 2003 mass table \cite{au03}.
We only include
nuclei for which the experimental error on the binding energy is 
less than 200 keV, and make a further selection on the experimental 
data to  avoid special binding effects not related to the mechanisms we have
discussed.  We drop nuclei with $N=Z$ because of their Wigner 
energy \cite{wig1, wig2} contribution.
We also drop odd-odd nuclei which have an additional
neutron-proton pairing \cite{mad87,fb07}.   Finally, to isolate isospin
dependencies, we require that the nuclei have positive isospin ($N>Z$).
The sizes of the resulting data sets are given in Table II.

For each of the data sets we fit the OES with the following 
fitting functions: 
\newcounter{bean}
\begin{list}{\roman{bean})}{\usecounter{bean}}
\item A constant $c_0$. This is the naive BCS form, and 
sets the scale for any improvements of the phenomenology.

\item The  one-term, two-parameter function
$c_\alpha A^{-\alpha}$ with prefactor $c_\alpha$ and exponent $\alpha$ which provides
the minimum rms deviation.  This shows the best one can do with 
two parameters, irrespective of justification.

\item The two-term expression $c_1 + c_2/A$. The constant $c_1$ 
represents the BCS correlation energy, assumed independent of $A$, 
and the  term $c_2/A^{-1}$ represents interaction and single-particle
contributions. We use additional subscripts $o,e$ and $n,p$ to distinguish
the different OES measures.

\end{list}

The rms residuals of each of the four OES's with each
of the three fit functions are shown
in Table II. 
We also show the fit parameters for fits i) and iii).

\begin{table}
\caption{Fits to the OES.  Column 2 reports the size of each data set.
Columns 3-5 show the rms residuals for the three fit functions discussed
in the text.  Columns 6-8 gives the fit parameters for models $i)$ and
$ii)$. Units are MeV. }
\begin{tabular}{|c|c|ccc||c|cc|}
\colrule
\multicolumn{2}{|c|}{Data set} & \multicolumn{3}{c||}{rms residual}
&\multicolumn{3}{c|}{fit parameters} \\
\colrule
OES  & Size &$_i$ &$_{ii}$ &$_{iii}$ &$c_0$ &$c_1$ &$c_2$ \\
\colrule
$\Delta^{(3)}_{on}$& 443 &  0.313  & 0.254 & 0.270 &1.04 & 0.82  & 24.   \\
$\Delta^{(3)}_{en}$& 442 &0.420  & 0.278  & 0.308 &1.32 & 0.94  & 41.   \\
$\Delta^{(3)}_{op}$& 418 & 0.275  & 0.220 & 0.231 &0.96 & 0.75  & 25.   \\
$\Delta^{(3)}_{ep}$&  407 &0.455  & 0.245 & 0.270 &1.64 & 1.11  & 59.   \\
\colrule
\end{tabular}
\end{table}

\subsection{Quality of fits}

Let us first examine the quality of the fits.  The simplest, just a 
constant, has an rms residual of about 0.3 MeV for the odd OES
and 0.45 for the even  OES measure.    The residuals
are decreased substantially, by up to 46\%, going to the power law fit,
model ii).    Model iii) with two terms
decreases the rms residual up to 41\% and are almost competitive with
model ii).  In view of its better theoretical justification, 
we would advocate using it instead of the power-law fit.
We note that the rms residual is larger for the
even OES than for the odd one.  This already suggests that 
shell fluctuations associated with the single-particle mechanism may be 
stronger than those affecting the time-odd contribution.

\subsection{Neutron fit parameters}

We now examine what information can be extracted from the fit parameters that
we have obtained in model iii).  
Examining the neutron OES,
we first note that the constant terms in the even and odd OES 
are close to each other;  the ratio of the difference to the sum is
$0.06/0.88\approx 7$\%.  
Since the BCS is the only $A$-independent contribution, and it is common
to both the even and the odd OES,  the near-equality is 
just what we expect.   Comparing
the values to the one-term constant $c_0\approx 1.0-1.3$, we see that
the assigned BCS pairing represents only 
$\sim 75\%$ of the total on average.

The coefficients $c_2$,  which we associate with the mean-field mechanisms,
are quite different in the even and odd OES.
That is also anticipated,  since the two mechanisms we discussed affect
them differently. From the coefficients $c_{2o},c_{2e}$ and
an assumed single-particle contribution according to Eq. (\ref{fg}), we can
extract 
average values for matrix elements in the time-odd contribution.  These 
values are:
\be
\label{to-results}
v_{i\bar i,i\bar i}\approx -49\,\,{\rm MeV;}\,\,\,\,\, {\bar v}\approx -30\, {\rm
MeV}
\ee 
The signs and relative magnitudes are as expected for a basic $NN$
interaction is attractive and short-ranged.  For such interactions,
the exchange contribution is large and attractive for $v_{i\bar i,i\bar i}$
but not for $\bar v$.  However, we cannot expect to extract reliable
values from this simple approach for reasons already mentioned.

It is useful at this point to 
average the even and odd OES to emphasize the BCS correlation contribution
and to look at the differences as well which will emphasize the
mean field contribution.  We define parameters 
$c_{k\pm} = (c_{ko}\pm c_{ke})/2$ and show the fits in this form in Table III.  
The entries for the averages show again the relative importance of 
the mechanisms.  The two terms give equal contribution at mass $A$ 
satisfying $c_{1+}=c_{2+}/A$, giving 
$A\sim 40$ as the dividing point between light nuclei
dominated by non-BCS pairing and heavier nuclei dominated by BCS
pairing.
We have already remarked that $c_{1-}$ is small, as expected
from a BCS pairing mechanism.  The $c_{k-}$ are 
essentially determined by the curvature of the binding energy
surface in the two directions, along the $Z$ axis for $c_{k-p}$ 
and the $N$ axis for $c_{k-n}$.  
In the liquid drop model, this curvature is mostly due to the 
Coulomb and symmetry terms in the formula.  The curvature can be
largely fit with the $c_{2-}$, as expected from the functional form
of the symmetry term, $\sim (N-Z)^2/A$.  The Coulomb is different,
however, as we discuss below.

\begin{table}[h!]
\caption{Averages and differences of the OES parameters,
$c_{k\pm}= (c_{ko}\pm c_{ke})/2$. 
Columns 2-3 from the fit $iii)$ shown in Table II.  In the last
row, the proton OES has been adjusted by subtracting the Coulomb
interaction as discussed in the text.  Units are MeV.
}
\begin{tabular}{|c|cc|}
\colrule
Parameter & $c_{1\pm}$ &$c_{2\pm}$ \\
\colrule
$c_{k+n}$& 0.88 & 33.    \\
$c_{k+p}$& 0.93 & 42.   \\
$c_{k-n}$&  0.06 & 8.    \\
$c_{k-p}$&  0.18  & 17.    \\
$c_{k-p}^*$& 0.10 & 14.    \\
\colrule
\end{tabular}
\end{table}

\subsection{Proton fit parameters}

We now turn to the proton OES.  We first note that the $c_{1+}$  has
a slightly larger value than is found for the neutrons.  It is
tempting to interpret this as
evidence for a stronger BCS pairing correlation energy for protons.
There is of course additional 
physics associated with the mean-field aspects of the Coulomb interaction, 
but that affects only the difference coefficient $c_{i-p}$.  From Table III we see that the 
$c_{i-}$ parameters are not identical for neutrons and protons.
The liquid-drop Coulomb energy,
proportional to $Z^2/A^{1/3}$, gives rise to contributions to 
$\Delta^{(3)}_{o p}$ or $\Delta^{(3)}_{ep}$ that have a very flat 
dependence on $A$.  Parametrizing
it according to model $iii)$ gives Coulomb contributions
$c_{i+p}\approx 0 $ and $c_{1-p} \approx 0.08$ MeV ; $c_{2-p}\approx 3.$
MeV. In the bottom row, we subtract this contribution from the proton
parameters.  One sees that $c_{1-p}$ is reduced to a value $c^*_{1-p}$ much
closer to the neutron value, $c_{1-n}$.

\section {Summary and conclusions}
The odd-even staggering of the binding energy
depends on several effects. While a simple inverse power dependence  can fit
the general trends with mass number $A$, theory suggests
different dependencies for the various terms. 
We have argued that three of the important contributions are
single-particle energy spacing, BCS pairing, and particle-particle 
interactions, and that their $A$-dependence is given by
$A^{-1}$ for the single-particle
energy contribution, $A^{0}$, i.e., constant, for
BCS pairing, and $A^{-1}$ for contributions from diagonal 
orbital interactions.
A fit to the OES obtained from experimental 
masses indicates a significant  improvement in the rms 
when both  $A^{-1}$ and constant terms are included. 
The rms values for  fits with these two-term forms are slightly larger than for
the values for the single inverse power form with an adjusted exponent. 
While both of  these involve two free parameters 
the latter has no obvious theoretical basis, while the former does.  

The results for the constant in the fit show that an $A$-independent 
BCS pairing requires on the order of 25\% contribution from 
single-particle
and time-odd interactions to the total average OES.  The part left, attributable
to BCS pairing, is slightly larger for protons than neutrons.  This
division implies that mean-field effects are more important than 
BCS pairing in light nuclei ($A<\sim 40$) and vice versa for heavy nuclei. Furthermore
the diagonal Coulomb interaction is visible in the extract constants
and is consistent with the liquid drop formula.  The $1/A$ contributions
can associated with the mechanisms of Table I, and a values for the
interaction energies can be extracted.  However, this part of the analysis
is more speculative, in view the additional assumptions that must be introduced.

\section*{Acknowledgment}

We acknowledge helpful discussions with W. Nazarewicz and P.-H.~Heenen.
This work was supported by the UNEDF SciDAC Collaboration under DOE
Grant DE-FC02-07ER41457 and by the Institute of Nuclear Theory under 
Grant DE-FC02-07ER41132.

\end{document}